# New measures for evaluating creativity in scientific publications

Simona Doboli, Hofstra University
Fanshu Zhao and Alex Doboli, Stony Brook University

## 1. INTRODUCTION

The main objective of our ongoing research work is understanding the group dynamics of the scientific creative process, including the flow of research ideas, the creation of new ideas, and their impact on future ideas in collaborative communities. The principal means for pursuing this objective is by analyzing the main outputs of research work, such as scientific publications and their citations. The list of citations referenced in a publication gives an indication of the main sources of information related to and used by the paper. But, by itself, arguably, it does not inform which citations influenced the main contributions of the paper and how. Our underlying assumption is that creative ideas are novel and useful combinations of old ideas or knowledge [Amabile 1983]. Thus, understanding which citations have a direct contribution to a paper's main novel ideas is critical as well as understanding which papers produce truly creative ideas (i.e. ideas that are both novel and useful).

In order to understand how new ideas form current ideas and then propagate through a research community, a novel analysis methodology of the creative contribution of each citation is needed. The number of citations is considered to be by and large the main way of measuring a paper's influence [Wanga et al. 2013]. More recently, citations have been used to gain insight about the dynamics of a research domain, including the temporal patterns in which papers are cited [Wanga et al. 2013], models to predict the total number of citations and the impact factor of journals [Wanga et al. 2013], the relation between creativity and co-occurence of citations [Uzzi et al. 2013], and the main path of a scientific domain's evolution [Lucio-Arias and Leydersdorff 2008].

However, a common characteristics of current work is that citations are used as the single, main source of information for modeling scientific impact in large communities. There is little meaning assigned to a citation, e.g., its purpose or context of use in a paper, even though this knowledge is important in assessing the transformation and evolution of creative ideas within a research community. Citations are usually not correlated with the content of the paper, hence it is difficult to distinguish between citations that give a general reference to a domain or problem (such citations can arguably be replaced by other references) and citations in which the referred idea is the starting (triggering) point in the construction of a new solution. Second, research work is likely to be cited multiple times over time by the same group. Such groups usually work on a similar problem and/or have transformed the original idea for new applications. This insight helps understanding the importance of the cited paper for the specific community as well as the cited idea's flexibility in being transformed to tackle new applications. Analyzing clusters of citations by the same research group (rather than individual citations) offers new insight about the usefulness of cited work and its importance in spawning new research.

This presentation discusses two possible metrics for estimating the creativity of a scientific publication. Citations are clustered depending on their originating group (e.g., research laboratory) and classified into seven categories based on their purpose. This information is then used to evaluate the uniqueness (novelty) and usefulness of the presented contribution with respect to related work within the related community. Our study refers to publications in the area of high performance analog circuit design, a modern area in technology.

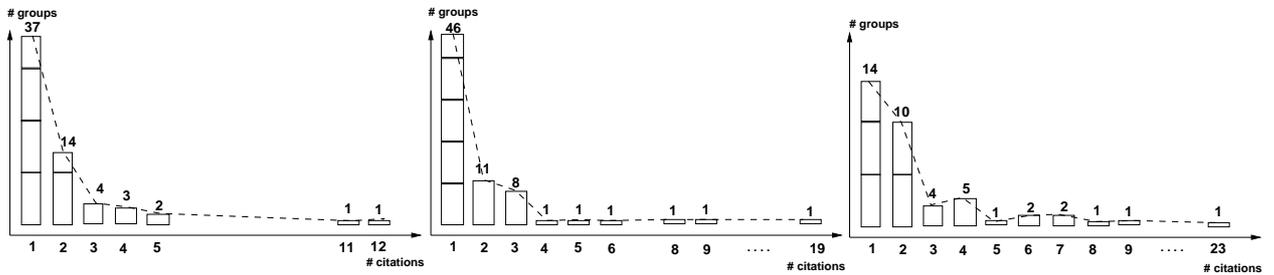

Fig. 1. Distributions of citations for three highly-cited papers

We think that devising metrics to estimate a paper's creativity is an important step towards a deeper understanding of the dynamics of the research process in scientific communities. For example, some important questions can be tackled, like the correlations between influence and creativity, the type of references in highly influential versus creative papers, the number and nature of ideas combined by influential versus creative papers, the importance and dynamics of self-citations, and many more.

## 2. CORRELATING CREATIVITY IN DESIGN WITH CREATIVITY IN SCIENTIFIC PUBLICATIONS

The research literature in design innovation presents various models to quantitatively evaluate the creativity of design solutions [Shah et al. 2003; Ferent and Doboli 2011; Umbarkar et al. 2014]. For example, Shah et al. [Shah et al. 2003] estimate the novelty index of a design feature as $S_j = \frac{T_j - C_j}{T_j} \times 10$. Variable $T_j$ is the number of designs in a set having feature $j$. Variable $C_j$ is the number of designs from the set using the same implementation of feature $j$. The novelty measure $MN_i = \sum_{j=1}^{n}(f_j \times S_j)$, where $n$ is the total number of features of the design, $f_j$ is the weight of feature $j$, and $S_j$ is the novelty index of feature $j$ [Shah et al. 2003]. The second measure, variety, is defined as in [Shah et al. 2003] $MV_i = 10 \times \sum_{j=1}^{m}(fj \times \frac{\sum_{k=1}^{n}(V_k \times b_k)}{MAX_V})$, where $n$ is the number of abstraction levels that can be identified for the solution set ($n$ abstraction levels spawn from the top, most conceptual description to the lowest, most detailed realization), $V_k$ is the variety index (weight) assigned to each level, and $b_k$ is the number of alternatives branching out from level $k$. $MAX_V$ is the populations maximum achievable variety, and equals the number of designs in the set times the variety index for the first level (the case when all circuits in the set implement different conceptual principles). We adapted the original metrics by Shah et al. for estimating the creativity of analog circuit designs [Ferent and Doboli 2011].

Before we adapt the above measures [Shah et al. 2003] to characterizing the novelty of a scientific publication, we gathered the following insight into the distribution of citations for different reasons. The two observations below relate a circuit design paper's creativity to its citation list:

*i. Citation roles.* We analyzed manually more than 30 papers on high-frequency analog circuit design. The papers were published since 2000 by arguably the most prestigious journals in this area (e.g., IEEE Journal of Solid State Electronics, IEEE Transactions on Circuits & Systems I, and IEEE Transactions on Circuits & Systems II). We looked at the reason and place of a citation, and classified it in one of the following seven categories: (a) general domain or problem reference, (b) reference to a particular design or method addressing the same problem or goal as the citing paper, (c) reference to a feature used in the paper's method or approach, (d) reference to using the design as a building (enabling) block in a more complex solution, (e) reference to an analysis method also used in the paper, (f) reference for an experimental comparison, and (g) reference to a design fact (property) established by the cited paper and used to justify the paper's assumptions. With this additional information a citation's influence in idea generation and propagation can be described and possibly measured more precisely. For example, a reference to a feature used in the paper can indicate that a citation has contributed to a paper's novel approach or idea, whereas a general reference can be a justification of the problem or an acknowledgement to existing approaches.

The classification information can be correlated to the attributes used in assessing creativity. Categories (b) and (f) express the uniqueness of the ideas proposed in a paper by contrasting them with similar work. The comparison highlights the nature of the design contribution including the abstraction level (i.e. conceptual level vs. physical implementation [Ferent and Doboli 2011]), the number of design variables, and the similarity of the work with other solutions. Categories (c), (d), (e), and (g) describe the usefulness of the presented ideas, such as support for devising new solutions by incorporating some design features, using specific analysis results (i.e. formula for performance parameters), or relying on certain conclusions as basic assumptions in the new work (e.g., the significance of specific design options). We think that citations in category (a) are less relevant for assessing creativity as they could be arguably replaced by citations with a similar role.

*ii. Citation clusters*. We observed that the number of references to a paper by different groups has a long-tail Levy like distribution. Figure 1 shows the distribution of citations for three, highly-cited papers. For example, the left most plot corresponds to the paper R. Castello, P. R. Gray, "A high-performance micropower switched-capacitor filter", IEEE Journal of Solid-State Circuits, Volume 20, Issue 6, December 1985, pp. 1122 - 1132, ISSN: 0018-9200. The paper received a total of 153 citations in IEEE and non-IEEE publications as well as in patents (according to IEEEXplore). The citations in IEEE and non-IEEE publications included 37 research groups which cited the paper a single time, 14 groups which cited it twice, four groups which cited it three times, and so on. One group cited the paper 11 times and another group 12 times. Similar trends can be observed for the other two plots too.

Some patterns were identified for the citation clusters in Figure 1. First, large citation clusters (i.e. 5 or more citations per cluster) correspond to three situations: (i) auto-citations (categories (d), (f), (g)), (ii) work which applied the cited ideas to new applications or problems (categories (d) and (e)), and (iii) work which discussed alternative solutions to similar problems (category (b)). Cases (i) and (ii) describe situations that define the usefulness of the concepts, while case (iii) presents situations that specify the uniqueness of the ideas. Second, small citation clusters (i.e. one or two citations per cluster) often indicate solutions that improved upon the cited design by incrementally modifying some of the design features but without altering the solution concept (category (c)). These situations express the flexibility of the cited design in being transformed to tackle a larger variety of design needs and constraints [Ferent et al. 2013]. Note however that few smaller clusters described new applications of the cited ideas (i.e. analysis of the circuits), hence the relation between the seven categories of citations and the size of the citation clusters needs to be further studied.

## 3. TWO METRICS FOR EXPRESSING CREATIVITY OF A PUBLICATION

By adapting the novelty measure [Shah et al. 2003], we propose that a publication's novelty is estimated using the following equation:

$$Novelty = \frac{1}{MAX_{C_{b \cup f}} \sum_{i \in C_{b \cup f}} \frac{f_i}{Cit_i}} \quad (1)$$

where $C_{b \cup f}$ is the set of papers in categories (b) and (f). $i$ represents a cluster of papers in $C_{b \cup f}$ authored by the same research group. $Cit_i$ is the size of the cluster (hence the number of citations offered by the group to the same paper), and $MAX_{C_{b \cup f}}$ is the maximum size of the cluster. $f_i$ is the weight associated to the differentiating features of the cited work and its alternative.

The rational behind the metric is as follows. Papers in categories (b) and (f) describe alternatives to the cited ideas, hence having more alternatives reduces uniqueness. The number of citations $Cit_i$ reflects the various specific differences discussed between an alternative approach and a solution assuming that each new paper by the same group focused on a single aspect. $MAX_{C_{b \cup f}}$ is a measure of the maximum conceptual difference between the cited idea and an alternative.

A publication usefulness is estimated by the following equation:

$$Usefulness = \frac{\sum_{i \in C_c} w_i Cit_i}{MAX_{C_c}} + \frac{\sum_{j \in C_d} w_j Cit_j}{MAX_{C_d}} + \frac{\sum_{k \in C_e} w_i Cit_k}{MAX_{C_e}} + \frac{\sum_{l \in C_g} w_l Cit_l}{MAX_{C_g}} \quad (2)$$

The meaning of the parameters is similar to equation (1). The usefulness of a paper increases with the number of research groups that cited it through citations in categories (c), (d), (e), and (g). The number of citations $Cit$ by the same group is divided by the maximum number of citations $MAX_C$ in the same category to reflect the varying degree of usefulness of the cited work for different groups (e.g., less citations implying a lower usefulness).

Our ongoing work focuses on validating the proposed metrics for a broad variety of publications in analog circuit design.